\date{}
\documentclass[showpacs,superscriptaddress]{revtex4}
\usepackage{graphicx,psfrag,amsmath,amssymb,amsfonts,bbm,latexsym,color,dcolumn,epsf}

\begin{document}
\title{Model for resonant photon creation in a cavity with time dependent conductivity}

\author{Mart\'\i n Crocce}
\affiliation{Department of Physics, New York University, 4 Washington Place, New York, New York 10003}

\author{Diego A. R. Dalvit}
\affiliation{Theoretical Division, MS B213 Los Alamos National Laboratory, Los Alamos, NM 87545}

\author{Fernando C. Lombardo}
\author{ Francisco D. Mazzitelli}
\affiliation{Departamento de F\'\i sica {\it Juan Jos\'e Giambiagi}, FCEyN UBA,
Facultad de Ciencias Exactas y Naturales, Ciudad Universitaria,
Pabell\' on I, 1428 Buenos Aires, Argentina}

\date{today}

\begin{abstract}
In an electromagnetic cavity, photons can be created from the
vacuum state by changing the cavity's properties with time. 
Using a simple model based on a massless scalar field,
we analyze  resonant photon creation induced by the
time-dependent conductivity of a thin semiconductor film contained
in the cavity. 
This time dependence  may be achieved by
irradiating periodically the film with short laser pulses. This
setup offers several experimental advantages over the case of
moving mirrors.
\end{abstract}

\pacs{03.70.+k;42.50.Lc;42.50.Nn}

\maketitle

\newcommand{\beq}{\begin{equation}}
\newcommand{\eeq}{\end{equation}}
\newcommand{\dalam}{\nabla^2-\partial_t^2}


\section{Introduction}

In quantum field theory time-dependent boundary conditions or
time-dependent background fields may induce particle creation, even when
the initial state of a quantum field is the vacuum \cite{birrel}.
In the context of quantum electrodynamics, uncharged mirrors in
accelerated motion can in principle create photons. This effect
is referred to in the literature as the dynamical Casimir effect,
or motion-induced radiation \cite{reviewdod}. In particular, when
the length of a high Q electromagnetic cavity oscillates with one
of its resonant frequencies, the number of photons inside the
cavity accumulates slowly and grows exponentially with time. Many
authors have considered this problem using different
approximations: from toy models of scalar fields in $1+1$
dimensions \cite{varios1+1} to the more realistic case of scalar
and electromagnetic fields in three dimensional cavities
\cite{us, others}. Arbitrary periodic motion of the boundary of an
ideal cavity has been studied in \cite{vcubo}.
The relevance of finite temperature effects  and
losses have also been considered \cite{plunien}.

Unlike the static Casimir effect \cite{mostep}, that has been
measured with increasing precision in the last years \cite{exp},
an experimental verification  of the dynamical counterpart is
still lacking.  The main reason is that typical resonance
frequencies for microwave cavities  are of the order of GHz.  It
is of course very difficult to make a mirror to oscillate at such
frequencies. Several alternative proposals have been investigated
in which the physical properties of the medium inside the cavity
change with time, but keeping fixed the boundary of the cavity.
Yablonovitch \cite{yablo} pointed out that a nonlinear optical
medium whose index of refraction changes rapidly with time
accelerates the vacuum fluctuations, and may create photons.
Nonlinear optics may be used to produce effective fast moving
mirrors.

Closer to our present work, Lozovik {\it et al.} \cite{lozo}
proposed that one could change effectively the length of a cavity
by irradiating with ultra-short laser pulses a thin semiconductor
film deposited on one of the walls of the cavity. They evaluated
the number of photons created using an  instantaneous
approximation in which the cavity suddenly changes its length.
Moreover, the conductivity of the thin film was assumed to vanish
before excitation and to be infinite after irradiation. Cirone
{\it et al.} \cite{cirones} considered the photon generation
produced by a time-dependent dielectric, using also an
instantaneous approximation.

From an experimental point of view, the idea of changing the
effective length of a cavity by irradiating a semiconductor is
promising \cite{carugno}. The theoretical aspects of the problem
are still not completely clear.  The main goal of this paper is to
analyze the proposal of Ref. \cite{lozo} with a more realistic
model that takes into account both the finite conductivity of the
semiconductor film and the characteristic times of excitation and
relaxation. We will also consider the possibility of periodic
oscillations of the conductivity of the semiconductor, as this
will enhance particle production for certain resonance
frequencies.

We  will therefore consider a rectangular cavity  of fixed size,
with a thin semiconductor film inside it. We will model the
electromagnetic field by a massless scalar field and the film
conductivity with a sharp potential. This model has been
considered in previous works to describe partially transmitting
mirrors \cite{barton}, and it will be enough for our purposes.
The modes for a massless scalar field inside such a cavity are
described in Section II. When the film is irradiated with a short
laser pulse its conductivity  rapidly increases. After a
relaxation  time the conductivity drops down to the initial
value. The irradiation-excitation-relaxation cycle can be repeated periodically.
Although the conductivity may change several orders of magnitude, when the
initial conductivity is high enough (as is the case for a
semiconductor, see below),  the time dependent frequency of the
modes varies periodically with time with a small amplitude. This
will allow us to use standard methods to compute the evolution of
the modes inside the cavity, and to show that for resonant
external frequencies the number of photons inside the cavity
grows exponentially with time. This evolution is described in
Section III. In Section IV we estimate the number of photons for
realistic values of the different parameters, and we also discuss
the experimental advantages of the proposal over the moving
mirror case.  Section V contains the conclusions of our work.


\section{The model}

We consider a three dimensional model of a massless scalar field
within a rectangular cavity with perfect conducting walls with
dimensions $L_x$, $L_y$, and $L_z$. At the mid point of the cavity
($x=L_x/2$) a thin film of semiconducting material is located. We
model the conductivity properties  of such material by a potential
$V(t)$: The ideal limit of perfect conductivity corresponds to $V
\rightarrow \infty$, and $V \rightarrow 0$ to a `transparent'
material. In our more realistic scenario this potential varies
between a minimum value, $V_0$, and a maximum $V_{\rm max}$. The
Lagrangian of the scalar field within the cavity is given by 
\beq
{\mathcal L}=\frac{1}{2} \partial_{\mu}\phi\partial^{\mu}\phi -
\frac{V(t)}{2} \delta(x-L_x/2) \phi^2, \label{themodel} 
\eeq 
where
$\delta(x)$ is the one-dimensional Dirac delta function.
The use of an infinitely thin film is justified as long as the width of the slab is much 
smaller than 
the wavelengths of the relevant electromagnetic modes in the cavity. 
The
corresponding Lagrange equation reads, \beq (\nabla^2 -
\partial_t^2) \phi = V(t) \delta(x-L_x/2)\phi.
\label{fieldequation} \eeq

For the sake of clarity we divide the cavity into two regions: region I ($0\leq x \leq L_x/2$)
and region II ($L_x/2 \leq x \leq L_x$). Perfect conductivity at the edges of the cavity
imposes the following boundary conditions
\begin{eqnarray}
&& \phi_{\rm I}(x=0,t)=\phi_{\rm II}(x=L_x,t)=0; \nonumber \\
&& \phi_{\rm I}(y=0,t)=\phi_{\rm II}(y=L_y,t)=0; \nonumber \\
&& \phi_{\rm I}(z=0,t)=\phi_{\rm II}(z=L_z,t)=0. \label{BC1}
\end{eqnarray}
The presence of the semiconducting film introduces a discontinuity in the spatial derivative
of the field
(in the $\hat x$ direction), while the field itself remains continuous,
\begin{eqnarray}
\phi_{\rm I}(x=L_x/2,t)&=&\phi_{\rm II}(x=L_x/2,t), \nonumber \\
\partial_x\phi_{\rm I}(x=L_x/2,t)-\partial_x \phi_{\rm II}(x=L_x/2,t)&=&-V(t) \phi(x=L_x/2,t).
\label{BC2}
\end{eqnarray}
This can be seen by integrating out the field equation
(Eq. (\ref{fieldequation})) in the neighborhood of the film. As we
will see,  there are two sets of solutions to
Eqs. (\ref{BC1},\ref{BC2}). One characterized by the functions
\begin{equation}
\varphi_{\bf m}({\bf x}) = \sqrt{\frac{2}{L_x}}  \sin\left(
\frac{2 \pi m_x x }{L_x} \right) \sqrt{\frac{2}{L_y}} \sin\left(
\frac{\pi m_y y}{L_y} \right) \sqrt{\frac{2}{L_z}}  \sin\left(
\frac{\pi m_z  z}{L_z} \right), \label{sol1}
\end{equation}
where ${\bf m}=(m_x, m_y, m_z)$ are positive integers, and the other by
\begin{equation}
\psi_{\bf m}({\bf x},t) = \sqrt{\frac{2}{L_x}}
\sin\left(k_{m_x}(t)\,x\right)\sqrt{\frac{2}{L_y}}
\sin\left(\frac{\pi m_y  y}{L_y}\right)\sqrt{\frac{2}{L_z}}
\sin\left(\frac{\pi m_z  z}{L_z}\right). \label{sol2}
\end{equation}
The function $\psi_{\bf m}$ depends on $t$ through  $k_{m_x}(t)$,
which is the $m_x$-th positive solution to the following
transcendental equation \beq 2 k_{m_x}
\tan^{-1}\left(\frac{k_{m_x}L_x}{2}\right)=- V(t) .
\label{trascendentalequation} \eeq To simplify the notation, in
what follows we will write $k_m$ instead of $k_{m_x}$.

Let us define
\begin{equation}
\Phi_{\bf m}({\bf x}) = \left\{   \begin{array}{ll}
                 \varphi_{\bf m}(x,y,z) & 0 \le x  \le L_x/2 \\
                 \varphi_{\bf m}(x-L_x,y,z) & L_x/2 \le x \le L_x
                      \end{array}
        \right.
\end{equation}
\begin{equation}
\Psi_{\bf m}({\bf x},t) = \left\{   \begin{array}{ll}
                 \psi_{\bf m}(x,y,z,t) & 0 \le x  \le L_x/2 \\
                 -\psi_{\bf m}(x-L_x,y,z,t) & L_x/2 \le x \le L_x
                      \end{array}
        \right.
\end{equation}
These functions satisfy the boundary conditions Eq. (\ref{BC1}) and
Eq. (\ref{BC2}), and the orthogonality relations $\left(\Phi_{\bf m},
\Phi_{\bf n} \right) = \delta_{\bf m,n}$, $\left( \Phi_{\bf m},
\Psi_{\bf n} \right) = 0$, and $\left(\Psi_{\bf m}, \Psi_{\bf n}
\right)  = \left[ 1 - \sin(k_m(t) L_x) / k_m(t) L_x \right]
\delta_{\bf m,n}$, where we have used the usual inner product in
the intervals $[0,L_i]$. It is also easy to show that $
\left(\Phi_{\bf m}, \frac{\partial \Psi_{\bf n}}{\partial k_n
}\right) = \left(\Phi_{\bf m}, \frac{\partial^2 \Psi_{\bf
n}}{\partial k^2_n}\right)=0$.

For $t\leq 0$ the slab is not irradiated, consequently $V$ is
independent of time and has the value $V_0$. The modes of the
quantum scalar field that satisfy the Klein Gordon equation
(\ref{fieldequation}) are
\begin{eqnarray}
v_{\bf m}({\bf x},t)&=&\frac{e^{-i  w_{\bf m} t}}{ \sqrt{ 2
w_{\bf m}}}\Phi_{\bf m}({\bf
x})\label{v}\\
u_{\bf m}({\bf x},t)&=&\frac{e^{-i \bar \omega_{\bf m} t}}{ \sqrt{
2\bar\omega_{\bf m}}}\Psi_{\bf m}({\bf x},0)\,\,\, , \label{u}
\end{eqnarray}
where $ w_{\bf m}^2 = \pi^2 \left[\left(\frac{2m_x}{L_x}\right)^2
+ \left(\frac{m_y}{L_y}\right)^2 +
\left(\frac{m_z}{L_z}\right)^2\right]$, $\bar\omega_{\bf
m}^2=(k_m^0)^2 + \left(\frac{\pi m_y}{L_y}\right)^2 +
\left(\frac{\pi m_z}{L_z}\right)^2$, and $k_m^0$ is the $m$-th
solution to Eq.(\ref{trascendentalequation}) for $V=V_0$. At
$t=0$ the potential starts to change in time and the set $\{ k_m
\}$ of the eigenfrequencies of the cavity acquires a time
dependence through Eq.(\ref{trascendentalequation}).

Using Eqs.(\ref{v}) and (\ref{u}) we expand the field operator $\phi$ as
\begin{equation}
\phi({\bf x},t) =
\sum_{\bf m} \left[a_{\bf m} v_{\bf m}({\bf x}, t) + b_{\bf m} u_{\bf m}({\bf x}, t)
+a^\dagger_{\bf m} v^*_{\bf m}({\bf x}, t) + b^\dagger_{\bf m} u^*_{\bf m}({\bf x}, t) \right],
\end{equation}
where $a_{\bf m}$ and  $b_{\bf m}$ are annihilation operators.
Notice that the solutions in Eq.(\ref{v}) correspond to modes
with a node at $x=L_x/2$ and therefore their dynamics is not
affected by the presence of the slab (if the initial particle number
given by $\langle a_{\bf m}^\dagger a_{\bf m}\rangle$ is zero, it will remain zero for all times).
These modes are also decoupled from the modes $u_{\bf m}$ thanks to the orthogonality 
conditions.
Therefore, we will only consider the evolution of the modes given in Eq.(\ref{u}).

For $t\geq 0$ we write the expansion of the field mode $u_{\bf s}$ as
\beq
u_{\bf s}({\bf x},t>0) = \sum_{\bf m} P_{\bf m}^{({\bf
s})}(t) \Psi_{\bf m}({\bf x},t) .
\label{singlefunction}
\eeq
Replacing this expression into $(\nabla^2 -
\partial^2_t) u_{\bf s}=0$  we find
\begin{equation}
\ddot{P}_{\bf n}^{({\bf s})} +  \omega_{\bf n}^2 (t) P_{\bf
n}^{({\bf s})} = - \sum_{\bf m} \left[ \left( 2 \dot{P}_{\bf
m}^{({\bf s})} \dot{k}_{m} + P_{\bf m}^{({\bf s})}
\ddot{k}_{m} \right)   g_{\bf m n}^{(A)} + P_{\bf
m}^{({\bf s})} \dot{k}_{m}^2  g_{\bf m n}^{(B)} \right],
\label{eqP}
\end{equation}
where $\omega_{\bf m}^2(t) = k_m^2(t) + \left(\frac{m_y
\pi}{L_y}\right)^2 + \left(\frac{m_z \pi}{L_z}\right)^2$. Note
that $\bar\omega_{\bf m} =\omega_{\bf m}(0)$.  The coefficients
$g_{\bf m n}^{(i)}$ read
\begin{eqnarray}
g_{\bf m n}^{(A)} &=&  \frac{1}{\left(\Psi_{\bf n}, \Psi_{\bf n}
\right) }
\,   \left(  \frac{\partial \Psi_{\bf m}}{\partial k_m} , \Psi_{\bf n} \right) , \nonumber \\
g_{\bf m n}^{(B)} &=&  \frac{1}{\left( \Psi_{\bf n}, \Psi_{\bf n}
\right)} \,  \left(  \frac{\partial^2 \Psi_{\bf m }}{\partial k_m^2}
, \Psi_{\bf n} \right). \label{coef12}
\end{eqnarray}
Imposition of continuity of $u_{\bf s}$ and $\dot u_{\bf s}$ at
$t=0$ gives the following initial conditions
\begin{eqnarray}
P^{({\bf s})}_{\bf m}(0) &=& \frac{1}{\sqrt{2\bar\omega_{\bf m}}} \delta_{\bf s,m}  \nonumber \\
\dot{P}^{({\bf s})}_{\bf m}(0) &=& -i\sqrt{\frac{\bar\omega_{\bf m}}{2}}
\delta_{\bf s,m}, 
\label{ic1}
\end{eqnarray}
provided that $V(t)$ and its time derivative are continuous at
$t=0$.


\section{Resonant photon creation}

We are interested in the number of photons created inside the
cavity. Hence we focus in resonance effects induced by periodic
oscillations in the conductivity $V(t)$, which translates into
effective  periodic changes in the modes of the scalar
field. Therefore we start by considering a time dependent
conductivity given by
\begin{equation}
V(t) = V_0 + \left(V_{\rm max} - V_0\right) f(t)\,\,\, ,
\label{vgen}
\end{equation}
where $f(t)$ is a periodic and non negative function,
$f(t)=f(t+T)\geq 0$, that vanishes at $t=0$ and attains its maximum
at $f(\tau_e)=1$. In each period, $f(t)$ describes the excitation
and relaxation of the semiconductor produced by the laser pulse.
Typically, the characteristic time of excitation $\tau_e$ is the
smallest time scale and satisfies $\tau_e\ll T$. We expand $f(t)$
in a Fourier series
\begin{equation}
f(t)=f_0  + \sum_{j=1}^{\infty}\left[ l_j \cos (j\Omega
t)+h_j\sin(j\Omega t)\right]= f_0 + \sum_{j=1}^{\infty} f_j \cos(j \Omega t + c_j) ,
\label{expand}
\end{equation}
where $\Omega=2\pi/T$. Since $\tau_e$ is the smallest time scale,
on general grounds we expect the first $T/\tau_e$ terms in the
above series to be relevant. We will present a particular
example in Section IV.

Under certain constraints, large changes in $V$  induce only small
variations in $k$ through the transcendental relation between $k$
and $V$ (see Eq. (\ref{trascendentalequation})). In this case, a
perturbative treatment is valid and a linearization of such
relation is appropriate. Accordingly we write \beq k_n(t)=k_n^0
(1+\epsilon_n f(t)) , \label{kdet} \eeq where $\epsilon_n$ is
obtained after replacing Eqs. (\ref{vgen},\ref{kdet}) into Eq.
(\ref{trascendentalequation}) and expanding it to first order in
$\epsilon_n$. The result is \beq \epsilon_n = \frac{V_{\rm
max}-V_0}{L_x (k_n^0)^2 + V_0 \left(1+\frac{V_0 L_x}{4}\right)}
.\label{epsilon} \eeq The restriction for the validity of the
perturbative treatment is $V_0 L_x \gg V_{\rm max} / V_0>1$. These
conditions are satisfied for realistic values of $L_x$, $V_0$, and
$V_{\rm max}$ (see Section IV).  It is worth noticing that we are
interested in low eigenfrequencies, for which $k(t)\sim {\mathcal
O}(L^{-1}_x)$. Nonetheless the perturbative treatment is also
valid for $k\sim {\mathcal O}(V)$.

In what follows we will only consider expressions to first order
in $\epsilon_n$. To analyze the
possibility of parametric resonance we write the spectrum given in
Eq. (\ref{kdet}) as,
\beq k_n(t) = \tilde{k}^{0}_n (1+\epsilon_n
(f-f_0)),
\eeq
where $\tilde{k}^0_n \equiv k^0_n (1+\epsilon_{\rm
n}f_0)$ is a `renormalized' frequency. The equation for the
coefficients $P^{({\bf s})}_{\bf m}(t)$ (Eq. (\ref{eqP})) can now
we written to first order in $\epsilon_n$ as,
\beq
\ddot{P}^{({\bf
s})}_{\bf n} + \tilde{\omega}^2_{\bf n}  P^{({\bf s})}_{\bf n} = -2
\epsilon_n (k_n^0)^2 (f-f_0) P^{({\bf s})}_{\bf n} - \sum_{\bf m}
\left[ 2 \dot{P}_{\bf m}^{({\bf s})}  \epsilon_m k^0_m  \dot f + \
P_{\bf m}^{({\bf s})}  \epsilon_m  k^0_m \ddot f \right]  g_{\bf m
n}^{(A)} +  {\mathcal O}(\epsilon^2),
\eeq
where $\tilde\omega_{\bf n}^2 = (\tilde k_n^0)^2 + (\pi n_y /L_y)^2 +
(\pi n_z /L_z)^2$. This equation describes a set of coupled
harmonic oscillators with periodic frequencies and couplings. It
is of the same form as the equations that describe the modes of a
scalar field in a three dimensional cavity with an oscillating
boundary, and can be solved  using  multiple scale analysis (MSA)
\cite{us}. A naive perturbative solution of previous equations
in powers of $\epsilon_n$ breaks down after a short amount of time (this happens
for particular values of the external frequency such that there is a resonant
coupling with eigenfrequencies of the cavity). In order
to find a solution valid for longer times, we introduce a second time scale $\tau_n = \epsilon_n t$
and expand $P_{\bf n}^{({\bf s})}$ to first order  as $P_{\bf n}^{({\bf s})}(t) =
P_{\bf n}^{({\bf s})(0)}(t,\tau_n) + \epsilon_n P_{\bf n}^{({\bf s}) (1)} (t, \tau_n) +
{\mathcal O}(\epsilon_n^2)$.

The derivatives with respect to the time scale $t$ read
\beq
\dot{P}_{\bf n}^{({\bf s})} = \partial_t P_{\bf n}^{({\bf s})(0)} +
\epsilon_n \left[\partial_{\tau_n}  P_{\bf n}^{({\bf s})(0)} + \partial_t
 P_{\bf n}^{({\bf s})(1)}\right],\eeq
\beq
\ddot{P}_{\bf n}^{({\bf s})} = \partial^2_t P_{\bf n}^{({\bf s})(0)} +
\epsilon_n \left[2 \partial^2_{{\tau_n} t}  P_{\bf n}^{({\bf s})(0)} + \partial^2_t
 P_{\bf n}^{({\bf s})(1)}\right].\eeq
For the zeroth order term we get the equation of a
harmonic oscillator \beq P_{\bf n}^{({\bf s}) (0)} = A_{\bf n}^{({\bf s})}(\tau_n)  e^{i
\tilde{\omega}_{\bf n} t} + B_{\bf n}^{({\bf s})}(\tau_n) e^{-i \tilde{\omega}_{\bf n} t} .
\eeq
From Eq. (\ref{ic1}) the initial conditions for $A_{\bf n}^{({\bf s})}$
and $B_{\bf n}^{({\bf s})}$ are
\begin{eqnarray}
A_{\bf n}^{({\bf s})}(\tau_n=0) & = & \frac{1}{\sqrt{2\bar\omega_{\bf n}}}
\left(1-\frac{\bar\omega_{\bf n}} {\tilde{\omega}_{\bf n}} \right)
\frac{\delta_{{\bf s},{\bf n}}}{2} \approx
\frac{1}{\sqrt{2 \bar\omega_{\bf n}}}  \frac{(k_n^0)^2}{2 {\bar \omega}_{\bf n}^2}
f_0 \epsilon_n \delta_{{\bf s},{\bf n}} ,
\nonumber \\
B_{\bf n}^{({\bf s})}(\tau_n=0) &= & \frac{1}{\sqrt{2\bar\omega_{\bf n}}}
\left(1+\frac{\bar\omega_{\bf n}}{\tilde{\omega}_{\bf n}}\right) \frac{\delta_{{\bf s},
{\bf n}}}{2} \approx
\frac{1}{\sqrt{2 \bar\omega_{\bf n}}}  \left(
1 - \frac{(k_n^0)^2}{2 {\bar \omega}_{\bf n}^2} f_0 \epsilon_n \right)
\delta_{{\bf s}, {\bf n}} .
\label{ic2}
\end{eqnarray}

The first order term $P_{\bf n}^{({\bf s}) (1)}$  satisfies
\begin{eqnarray}
\partial^2_t  P^{({\bf s})(1)}_{\bf n} + \tilde{\omega}^2_{\bf n} P^{({\bf s})(1)}_{\bf n} =
-2 \partial_{t {\tau}_n}^2 P^{({\bf s})(0)}_{\bf n}  -2 (k_n^0)^2 (f-f_0) P^{({\bf s})(0)}_{\bf n} -
\sum_{\bf m}  \frac{\epsilon_m}{\epsilon_n} g_{{\bf m} {\bf n}}^{(A)}
\left[
2 \partial_t P_{\bf m}^{({\bf s})(0)}   k^0_m  \dot f +  P_{\bf m}^{({\bf s})(0)} k^0_m \ddot f
\right] ,
\end{eqnarray}
where $g_{{\bf m n}}^{(A)}$ is calculated to zeroth order in
$\epsilon$. The basic idea of MSA is to impose the condition that
any term in the right-hand side of the previous equation with a
time dependency of the form $e^{\pm i \tilde{\omega}_{\bf n} t}$ must
vanish. If not, these terms would be in resonance with the
left-hand side term and secularities would appear. Applying this
procedure we obtain
\begin{eqnarray}
\frac{1}{2}  \frac{d A_{\bf n}^{({\bf s})}}{d {\tau}_n}&=&  
\sum_{j} f_j
 \left\{- \frac{(k^0_n)^2}{4 i \tilde{\omega}_{\bf n}} B_{\bf n}^{({\bf s})}
e^{i c_j} \delta(2\tilde{\omega}_{\bf n}-\Omega_j) -  
\sum_{\bf m}  \frac{\epsilon_m}{\epsilon_n} \frac{\Omega_j}{4 i \tilde{\omega}_{\bf n}}
g_{{\bf m n}}^{(A)}\,k^0_m\,
[ (-\frac{\Omega_j}{2} - \tilde{\omega}_{\bf m}) A_{\bf m}^{({\bf s})} e^{i c_j}
\delta(\tilde{\omega}_{\bf n}
-\tilde{\omega}_{\bf m} - \Omega_j) + \right. \nonumber \\
&&
\left.
(-\frac{\Omega_j}{2}+\tilde{\omega}_{\bf m}) A_{\bf m}^{({\bf s})} e^{-i c_j} \delta(\tilde{\omega}_{\bf n}
-\tilde{\omega}_{\bf m}+\Omega_j) +
(-\frac{\Omega_j}{2}+\tilde{\omega}_{\bf m}) B_{\bf m}^{({\bf s})} e^{i c_j}
\delta(\tilde{\omega}_{\bf n}+\tilde{\omega}_{\bf m} -\Omega_j)]
\right\} ,
 \label{eqA}
\end{eqnarray}
and
\begin{eqnarray}
\frac{1}{2} \frac{d B_{\bf n}^{({\bf s})}}{d {\tau}_n} &=& 
\sum_j f_j
\left\{
\frac{(k^0_n)^2}{4 i \tilde{\omega}_{\bf n}} A_{\bf n}^{({\bf s})} e^{-i c_j}
\delta(2 \tilde{\omega}_{\bf n}-\Omega_j) + 
\sum_{\bf m} \frac{\epsilon_m}{\epsilon_n} \frac{\Omega_j}{4 i \tilde{\omega}_{\bf n}}
g_{{\bf m n}}^{(A)}\,k^0_m\, 
[(-\frac{\Omega_j}{2}-\tilde{\omega}_{\bf m}) B_{\bf m}^{({\bf s})} e^{-i c_j}
\delta(\tilde{\omega}_{\bf n}-\tilde{\omega}_{\bf m} - \Omega_j) + \right. \nonumber \\
&&
\left.
(-\frac{\Omega_j}{2}+\tilde{\omega}_{\bf m}) B_{\bf m}^{({\bf s})} e^{i c_j} \delta(\tilde{\omega}_{\bf n}
-\tilde{\omega}_{\bf m}+\Omega_j) +
(-\frac{\Omega_j}{2} +\tilde{\omega}_{\bf m}) A_{\bf m}^{({\bf s})} e^{-i c_j} \delta(\tilde{\omega}_{\bf n}
+\tilde{\omega}_{\bf m} - \Omega_j)] \right\} .
\label{eqB}
\end{eqnarray}
The coupling coefficients as defined in Eq.(\ref{coef12}) can be
computed using Eqs.(\ref{sol2},\ref{trascendentalequation}) and
(\ref{kdet}). However, we will not need the explicit expression in
what follows.

Since the function $f(t)$ contains the frequencies
$\Omega_j=j\Omega$, we expect resonant behavior and/or mode
coupling when one of the following relations are
satisfied
\begin{eqnarray}
\Omega_j &=& 2\tilde\omega_{\bf n} ; \nonumber\\
\Omega_j&=& | \tilde \omega_{\bf n}\pm \tilde\omega_{\bf m}| ,
\label{cond}
\end{eqnarray}
for any pair of modes ${\bf m}$ and ${\bf n}$.  As the
eigenfrequencies $\tilde\omega_{\bf n}$ are not equidistant, if
some of the conditions in Eq. (\ref{cond}) above are satisfied
by  a given set $(j,{\bf m},{\bf n})$, in general they will not
be satisfied for a different set $(j',{\bf m}',{\bf n'})$. This
can be checked by inspection in each particular case. Assuming
that this is the case, we can restrict the analysis to a single
Fourier mode in the expansion (\ref{expand}), i.e. $f(t)=f_0+f_j
\cos (j\Omega t+c_j)$.

Let us consider the most important case of parametric resonance,
i.e. when the external frequency is tuned with some of the
eigenfrequencies of the cavity, $\Omega_j=2\tilde{\omega}_{\bf
n}$. The contributions proportional to  $g_{{\bf m,n}}^{(A)}$ would be
different from zero if there exists an integer $j'$ such that
$(2\frac{j'}{j}\pm 1)\tilde\omega_{\bf n}=\tilde\omega_{\bf m}$. This may
happen only for some particular values of $L_x, L_y, L_z$. We will
assume that the condition is not satisfied. Therefore, as
mentioned above, we can restrict the analysis to the single $j$ Fourier
mode. Moreover, the resonant mode $\bf n$ is not coupled and we
are left with
\begin{eqnarray}
\frac{d A^{({\bf s})}_{\bf n}}{d {\tau}_n} = i\frac{(k^0_n)^2 f_j e^{i c_j}}{\Omega_j}
B^{({\bf s})}_{\bf n}  &;&
\frac{d B^{({\bf s})}_{\bf n}}{d {\tau}_n} = -i\frac{(k^0_n)^2 f_j e^{-i c_j}}{\Omega_j}
A^{({\bf s})}_{\bf n} .
\end{eqnarray}
The solution that satisfies the initial conditions Eqs. (\ref{ic2}) is
\begin{eqnarray}
A^{({\bf s})}_{\bf n}(\tau_n) &=& \frac{\delta_{{\bf s,n}}}{\sqrt{8 \bar\omega_{\bf n}}}
\left[
    \left(
    1-\frac{\bar\omega_{\bf n}}{\tilde\omega_{\bf n}}\right)
\cosh \left(\frac{(k^0_n)^2 f_j}{\Omega_j} \tau_n
    \right)  +
    \left(
    1+\frac{\bar\omega_{\bf n}}{\tilde\omega_{\bf n}}\right)  i e^{i c_j}
\sinh \left(\frac{(k^0_n)^2 f_j}{\Omega_j} \tau_n
    \right)
\right]\nonumber \\
 &\approx &  \frac{i e^{ic_j}\delta_{{\bf s,n}}}{\sqrt{2 \bar\omega_{\bf n}}}
\sinh \left(\frac{(k^0_n)^2 f_j}{\Omega_j} \tau_n \right)
    ; \nonumber \\
B^{({\bf s})}_{\bf n}(\tau_n) &=&  \frac{\delta_{{\bf s,n}}}{\sqrt{8 \bar\omega_n}}
\left[
    \left(
    1+\frac{\bar\omega_{\bf n}}{\tilde\omega_{\bf n}}\right)
\cosh \left(\frac{(k^0_n)^2 f_j}{\Omega_j} \tau_n
    \right)  -
    \left(
    1-\frac{\bar\omega_{\bf n}}{\tilde\omega_{\bf n}}\right)  i e^{-i c_j}
\sinh \left(\frac{(k^0_n)^2 f_j}{\Omega_j}
\tau_n
    \right)
\right] \nonumber \\
&\approx &
\frac{\delta_{{\bf s,n}}}{\sqrt{2 \bar\omega_{\bf n}}}
\cosh \left(\frac{(k^0_n)^2 f_j}{\Omega_j} \tau_n \right) .
\label{parametricsolution}
\end{eqnarray}
The corresponding number of created photons  with frequency
$\tilde{\omega}_{\bf n}=\Omega_j/2$  is given by
\begin{equation}
\langle {\mathcal N}_{\bf n}(t) \rangle = \langle b^\dagger_{\bf n} b_{\bf n}\rangle
=\sum_{\bf s} 2 \bar\omega_{\bf n} |A_{\bf n}^{({\bf s})} (t)|^2 \approx
\sinh^2 \left(\frac{(k^0_n)^2 f_j}{\Omega_j}  \epsilon_n t \right),
\label{photonnumber}
\end{equation}
that leads to an exponential growth in the number of photons in the mode with frequency
$\tilde{\omega}_{\bf n}$ at a rate $r_{\rm cond} = 2 (k_n^0)^2 f_j \epsilon_n/ \Omega_j$.
It can be shown that no other off-resonant frequencies $\Omega_{j'} \neq \Omega_j$ present in the
Fourier spectrum of $f(t)$ contribute to the exponential growth of the number of photons: in our model,
the photon frequency spectrum ${\mathcal F}(\omega)$ develops an exponentially growing
component at a single frequency $\omega=\Omega_j /2$.


\section{Numerical estimations}

To determine the typical range of values of the conductivity $V$
before and after the short laser pulse is applied to the
semiconductor we use the relation $V=4 \pi n_s e^2/m$, where $n_s$
is the surface charge density of free carriers in the
semiconducting film, $e$ is the electron charge, and $m$ is the
effective mass of the free carriers.  This relation was derived in
\cite{barton}, where it was shown that  our model Eq.
(\ref{themodel}) is equivalent to a plane-polarized
electromagnetic field propagating normally to an infinitesimally
thin jellium-type plasma sheet. When a laser field suddenly
impinges on the sheet, it produces time-dependent changes in the
surface charge density $n_s(t)$ that induce a time variation in
the conductivity $V(t)$ in our model. When the semiconductor is
illuminated in the GHz range it becomes an excellent conductor
(its conductivity being within $2 \%$ error from that of copper
\cite{private}). Therefore, using known values for the
conductivities of good conductors, we can fix $V_{\rm max} =
10^{16} {\rm m}^{-1}$.  When the laser field is not applied  we
can set $V_0 = 10^{10} {\rm m}^{-1} - 10^{13} {\rm m}^{-1}$, the
range of values for different semiconductors. For a cavity of size
$L_x\simeq 10^{-2} {\rm m}$,  and when the ratio between the maximum and
minimum conductivities is in the range $10^3 \leq V_{\rm
max}/{V_0} \leq 10^6$, we obtain from Eq.(\ref{epsilon}) the
following range of values of $\epsilon$, $10^{-8} \leq \epsilon
\leq 10^{-2}$. Therefore, these small values give {\it a
posteriori} justification for the perturbative approach we have
used. Note that $\epsilon$ is small even if the conductivity of
the film changes six orders of magnitude. For larger changes of
the conductivity our perturbative approach is not adequate.

\begin{figure}[t]
\includegraphics[width=8cm]{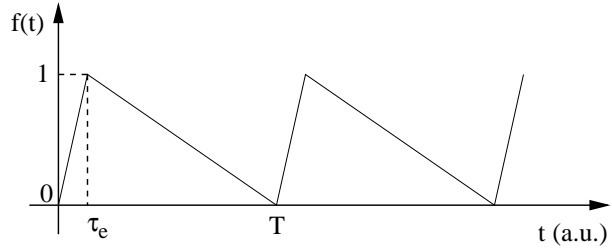}
\caption{Periodic excitation and relaxation of the semiconductor slab via 
ultra short laser pulses. The function $f(t)$ is proportional to the 
conductivity of the semiconductor (Eq. (\ref{vgen})), and mimics its evolution in time.}
\label{figure1}
\end{figure}

As we mentioned in the previous section, we consider a
periodic function $f(t)$ with period $T$.  To mimic the excitation and
relaxation processes, we choose the following linear ramps
\begin{equation}
f(t) = \left\{\begin{matrix} \frac{t}{\tau_e}&\mbox{for} ~ 0 < t \le
\tau_e \cr \frac{(T - t)}{(T - \tau_e)}& \mbox{for}  ~  \tau_e < t
\le T \cr
\end{matrix}
\right. 
\label{ramp}
\end{equation}
where $\tau_e$ is the excitation time scale, $\tau_e \ll T$.
This simple choice will be enough to estimate
the amplitude of the resonance effect. Expanding this $f(t)$ as
in Eq.(\ref{expand}) we obtain
$l_j  = f_j \cos c_j =  T [\cos(2 \pi j \tau_e/T)-1] / [ 2 \pi^2 j^2 \tau_e (1-\tau_e/T)]$, and
$h_j = - f_j \sin c_j = T \sin(2 \pi j \tau_e/T) / [ 2 \pi^2 j^2 \tau_e (1-\tau_e/T)]$, so that
\begin{equation}
f_j =  \frac{1}{\pi j (1-\tau_e/T)} \; \left| \frac{\sin(\pi j
\tau_e/T)}{\pi j \tau_e/T}\right| . \label{fj}
\end{equation}
On the one hand, in the limit $j \pi \tau_e /T \ll 1$ we get 
$f_j\approx 1 / \pi j$. This limit also corresponds to the instantaneous
approximation $\tau_e \rightarrow 0$,
in which the conductivity of the semiconductor slab suddenly changes
from low to high values. As mentioned in the introduction, a similar
instantaneous
approximation was used in \cite{lozo} and \cite{cirones}; however, our approach is more
general, not only because we  consider finite conductivities of the
slab, but also
because we analyze periodic excitations via the laser pulses.  On the
other hand, for large values of $j$ the coefficients $f_j$ decay as
$T / \tau_e j^2 \pi^2$. We have computed the Fourier
coefficients $f_j$ for the particular function $f(t)$ given in Eq.
(\ref{ramp}). However, provided that $1 \ll j \ll T/ \pi \tau_e$, it can be shown that
the result $f_j \sim 1/ \pi j$ is valid for any function $f(t)$ that grows
on a time scale $\tau_e$ and relaxes on a time $T$.

The number of created photons grows exponentially at a rate
$r_{\rm cond} =  2 (k_n^0)^2 f_j \epsilon_n / \Omega_j$.  When
$L_y, L_z \gg L_x$ we have $k_n^0 \approx  {\tilde \omega}_{\bf n} = \Omega_j /2$,
so that the rate reads  $r_{\rm cond} = \Omega_j f_j \epsilon_n/2$.
In the limit $\Omega_j \tau_e \ll 1$ this rate is independent of $j$ and is given
 by  $r_{\rm cond} = \epsilon_n/T$. In the limit $\Omega_j \tau_e \gg 1$ it
becomes  $r_{\rm cond} =2 (\Omega_j \tau_e)^{-1}\epsilon_n/T$.
In order to have resonant photon production for a
cavity of $L_x =10^{-2} {\rm m}$, the resonant frequency $2 \pi
j / T$ must be in the GHz range. With the present technology, it is possible to
 have femtosecond laser pulses with a repetition frequency up to 100 MHz,
 thus the resonance could be achieved by rather low values of $j$, 
in the range $1-10$.
The rate at which GHz photons would be produced depends now on whether the 
excitation time satisfies $\tau_e \ll 10^{-9} {\rm s}$ or $\tau_e \gg 10^{-9} {\rm s}$. Nowadays it
 is possible to reach values for $\tau_e$ as small as $10^{-12} {\rm s}$
 \cite{private}, thus yielding an estimated rate of $r_{\rm cond}\sim 10^8 \,\epsilon_n$
 Hz. Notice that this estimation remains true even for
 excitation times three orders of magnitude larger than $10^{-12}{\rm s}$, after
 which there is a suppresion factor of order $1/\Omega_j\tau_e$. Also notice
 that $r_{\rm cond}$  is acceptably large in all the range where $\epsilon$ 
was assumed to vary, namely $10^{-8}-10^{-2}$.

This estimation for the rate $r_{\rm cond}$ should be contrasted with that corresponding to
the moving mirror case. In the latter case the typical rate is
$r_{\rm mov}\approx \epsilon_{\rm mov} / T_{\rm mov}$, where
$\epsilon_{\rm mov}$ is the relative amplitude of the oscillations
of the mirror, and $T_{\rm mov}$ is the period of oscillation of
the moving mirror, which must be of order $10^{-9} {\rm s}$ for a
microwave cavity. If the oscillations are produced by deformations
of the surface, the value of $\epsilon_{\rm mov}$ cannot exceed $3
\times 10^{-8}$ \cite{reviewdod}. Then the accumulation of photons
is a very slow process, which requires extremely high values for
the $Q$-factor of the cavity. The ratio between the rate for
photon creation for the time-dependent conductivity model and that
of the moving mirror is
\begin{equation} \frac{r_{\rm
cond}}{r_{\rm mov}} \approx \frac{\epsilon_{n}}{\epsilon_{\rm
mov}}\frac {T_{\rm mov}}{T}\,\,\, ,
\end{equation}
which is of order $10^6 T_{\rm mov}/T$ for  $\epsilon_{n}\approx
10^{-2}$ and $\epsilon_{\rm mov}\approx 10^{-8}$. Even assuming
that it is possible to produce oscillations of the mirror with a
GHz frequency, and that the pulsing cycle of the laser is of some
MHz, this ratio is much larger than one.

Finally, in order to have resonance effects, the external
frequency should be tuned with the frequency of the resonant mode
with a high accuracy. In the moving mirror case, it has been shown
\cite{us} that if the external frequency is slightly detuned by an
amount $\Delta  \Omega_{\rm mov}$ from the resonant frequency
$\Omega_{\rm mov} = 2 \pi / T_{\rm mov}$, then the number of
photons grows exponentially as long as $\Delta\Omega_{\rm mov} /
\Omega_{\rm mov} <\epsilon_{\rm mov}$. A similar analysis shows
that,  in the setup described in this paper, an analogous
restriction applies, with $\epsilon_{\rm mov}$ replaced by
$\epsilon_n$. As we have mentioned above, $\epsilon_n$ can take
values much larger than $\epsilon_{\rm mov}$. Therefore, while the
fine tuning is severe in the moving mirror case, it is not so
critical for the case of a time dependent conductivity.

The discussion above shows that the present setup is much more
promising for the experimental verification of the dynamical
Casimir effect than the moving mirror one: the time dependent
conductivity may be achieved with the present technology, the rate
of photon creation is much larger, and the fine tuning is not so
severe.

\section{Conclusions}

In this paper we have studied a simple scalar model to mimic
photon creation induced by time-dependent changes in the
conductivity of a thin semiconductor that is periodically excited
by short laser pulses.  When the conductivity of the
semiconductor, placed inside a cavity of linear size in the ${\rm
cm}$ range, changes up to six orders of magnitude during the
excitation process induced by the laser pulses, the dynamical
equations for the field modes have time dependent coefficients
that oscillate in time with small amplitudes. This allowed us to
solve the equations for the modes using multiple scale analysis,
and to show that for some resonant excitation frequencies real
photons are created inside the cavity, their number growing
exponentially in time. Due to the very short excitation time of
the semiconductor ($\tau_e/T\ll 1$), it should be possible to tune
a cavity mode with the frequency of a high $j$ Fourier harmonic of
the time-dependent conductivity $V(t)$ of the semiconductor.
Remarkably, as long as $j \pi \tau_e/T \ll 1$, the rate of photon
creation is independent of $j$. It is then possible to produce
resonant effects with ultra-short laser pulses whose pulsing
frequency is well below the GHz range. We have also shown that
this setup offers several advantages over the case of
motion-induced radiation arising from moving mirrors, such as much
faster photo-production rates and milder fine tuning problems.

Several issues remain to be investigated, such as how to extend this research to the
more realistic full electromagnetic case, which involves 
Dirichlet and Neumann boundary conditions.
Based on previous results for the moving mirror case \cite{us}, we expect
an exponential growth of the number of created photons with
different rates for transverse electric and transverse magnetic modes.
Moreover, one should consider
the macroscopic properties of the semiconductor slab, such as its frequency dependent conductivity 
and permittivity \cite{nos2}. 

Another relevant aspect to investigate is how to disentangle Casimir 
photons from those that may be radiated from accelerated charges
and thermal fluctuations of micro-currents in the semiconductor film, 
as this is being irradiated by the external laser pulse. 

\section{Acknowledgments}
F. C. L. and F. D. M. were supported by Universidad de Buenos Aires, CONICET,
Fundaci\'on Antorchas and Agencia Nacional de Promoci\'on
Cient\'\i fica y Tecnol\'ogica, Argentina. We would like to thank N. Bonadeo and 
G. Carugno for useful conversations.

\end{document}